# Evolutionary crystal structure prediction and novel high-pressure phases.

Artem R. Oganov[1,2*], Yanming Ma[3], Andriy O. Lyakhov[1], Mario Valle[4], Carlo Gatti[5]

[1] *Deptartment of Geosciences, Department of Physics and Astronomy, and New York Center for Computational Sciences, State University of New York, Stony Brook, NY 11794-2100, USA.*
[2] *Geology Department, Moscow State University, 119992 Moscow, Russia.*
[3] *National Lab of Superhard Materials, Jilin University, Changchun 130012, P. R. China*
[4] *Data Analysis and Visualization Group, Swiss National Supercomputing Centre (CSCS), Cantonale Galleria 2, 6928 Manno, Switzerland*
[5] *CNR-ISTM, Istituto di Scienze e Tecnologie Molecolari, via Golgi 19, 20133 Milano, Italy*

Prediction of stable crystal structures at given pressure-temperature conditions, based only on the knowledge of the chemical composition, is a central problem of condensed matter physics. This extremely challenging problem is often termed "crystal structure prediction problem", and recently developed evolutionary algorithm USPEX (Universal Structure Predictor: Evolutionary Xtallography) made an important progress in solving it, enabling efficient and reliable prediction of structures with up to ~40 atoms in the unit cell using *ab initio* methods. Here we review this methodology, as well as recent progress in analyzing energy landscape of solids (which also helps to analyze results of USPEX runs). We show several recent applications – (1) prediction of new high-pressure phases of $CaCO_3$, (2) search for the structure of the polymeric phase of $CO_2$ ("phase V"), (3) high-pressure phases of oxygen, (4) exploration of possible stable compounds in the Xe-C system at high pressures, (5) exotic high-pressure phases of elements boron and sodium.

## I. Introduction

Crystal structure prediction problem occupies a central place in materials design. Solving this problem would also open new ways also for understanding the behaviour of materials at extreme conditions, where experiments are difficult (in some cases, prohibitively difficult).

Often the approach has been to compare the free energies of a number of candidate structures (usually taken from analogous materials, or constructed by chemical intuition). Data mining [2] efficiently explores databases of known crystal structures and, using correlations between structures adopted by different compounds, indicates a list of likely candidate structures. Problems arise almost every time when a totally unexpected and hitherto unknown structure is stable. A number of simpler intuitive empirical schemes (e.g. structure diagrams, polyhedral clusters – see [3]) have appeared in literature, but their application usually requires a large experimental data set or good understanding of the compound at hand.

Thanks to recent methodological developments, reliable structure prediction can be performed without any prior knowledge or assumptions about the system. Simulated annealing [4-7], minima hopping [8] and metadynamics [9-11] have been used with some success. For small systems, even relaxing randomly produced structures can deliver the stable structure [12]. Here we review the evolutionary algorithm USPEX (Universal Structure Predictor: Evolutionary Xtallography) [13-15] and a small selection of the results it has provided so far. Section II presents basics of the method, section III shows several interesting test cases (mostly on systems with a known ground state), while a number of applications to systems where the stable structure is unknown are presented in section IV. This review is an updated version of the previous account of the methodology [76].

## II. Evolutionary algorithm USPEX

Several groups attempted the pioneering use of evolutionary algorithms to structure prediction - for crystals [23-26], colloids [27] and clusters [28]. The algorithm developed by Deaven and Ho [28] is perhaps especially interesting as some of its features (real-space representation of structures, local optimization and spatial heredity) are similar to the USPEX method. Their algorithm has successfully reproduced the structure of the $C_{60}$ buckminsterfullerene, but has never been extended to heteroatomic clusters, nor to periodic systems (i.e. crystals). The algorithm of Bush and Woodley [23-25] was originally developed for crystals and successfully produced a starting model for solving the structure of $Li_3RuO_4$ [23]. However, subsequent systematic tests [24-25] showed frequent failures even for rather simple systems containing ~10 atoms/cell. Other drawbacks are that this algorithm requires experimental lattice parameters and simulations are very expensive, unless a cheap and crude heuristic expression is used for fitness. Unlike the Deaven-Ho algorithm and USPEX, in this method structures are represented by binary "0/1" strings, there is no local optimization and no spatial heredity.

In USPEX, structures are represented by fractional coordinates for the atoms and lattice vectors. USPEX operates with populations of structures; from them, parent structures are selected. The fitness of structures is the relevant thermodynamic potential derived from *ab initio* total energy calculations. The worst structures of a population are discarded; for the remaining structures the probability of being selected as parent is a function (e.g. linear) of its fitness rank. A new candidate structure is produced from parent structures using one of three operators: (i) heredity, which combines spatially coherent slabs (in terms of fractional coordinates) of two parent structures, while the lattice vectors matrices are weighted averages of the two parent lattice vectors matrices, (ii) permutation (as in [24-25]), which swaps chemical identities in randomly selected pairs of unlike atoms, (iii) lattice mutation, which distorts the cell shape by applying a random symmetric strain matrix. To avoid pathological lattices, all newly produced structures are rescaled to produce a predefined unit cell volume (a reasonable starting value should be supplied in the input, and then allowed to evolve during the run). Heredity enables very broad searches, while preserving already found local fragments of good structures, and introduces ideas of "two-phase" simulations. Permutation facilitates finding the optimal ordering of the atoms; in some situations (for systems with a large range in degree of chemical similarity between different atom types) it may be useful to swap only chemically more similar atoms (e.g. Al-Si in aluminosilicates). Lattice mutation enables better exploration of the neighbourhood of parent structures, prevents premature convergence of the lattice, and essentially incorporates the ideas of metadynamics in our search. The action of these variation operators is illustrated in Figs. 1,2.

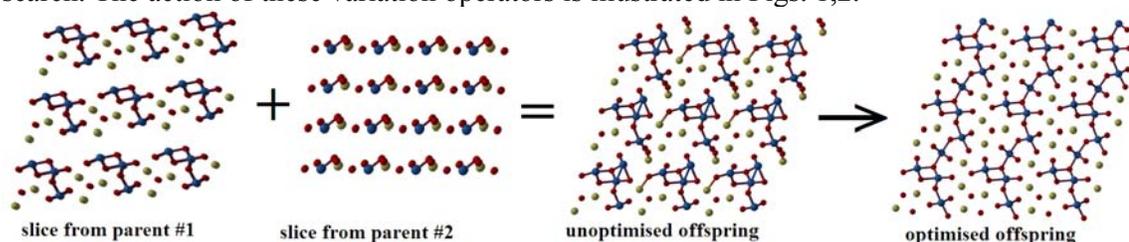

**Fig. 1. Heredity operator: slices of two parent structures, and the offspring structure before and after local optimization.**

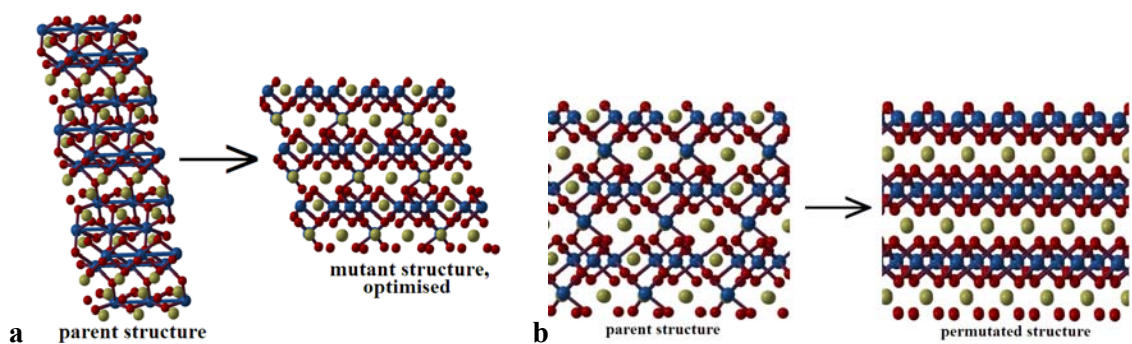

**Fig. 2.** Illustrations of lattice mutation and permutation operators.

Before new candidate structures are relaxed, they are tested against three constraints – first, all interatomic distances must be above the specified minimal values; second, cell angles must be between 60º and 120º; third, all cell lengths must be larger than a specified value (e.g. diameter of the largest atom). These constraints help to ensure stability of energy calculations and local optimization, and remove only redundant and infeasible regions of configuration space - thus the search is physically unconstrained. If in violation of these constraints, the candidate structure is discarded; otherwise, it is locally optimized (relaxed). Structure relaxations and energy calculations are done by external codes (currently, USPEX is interfaced with VASP [29], SIESTA [30], GULP [31]).

The relaxed structures are recorded and used for producing the next generation of structures. A new population of structures is made to contain one or more lowest-enthalpy structures from the previous population and the new structures produced using variation operators. Generation by generation, the above procedure is repeated in a loop.

The first generation usually consists of random structures, but it is possible to include user-specified structures. If lattice parameters are known, runs can be done in the fixed cell, but this is not required and in most cases simulations are done with variable cell shape. We have also improved [19] the algorithm by more exhaustive removal of lattice redundancies. For more details on the USPEX method, see [14,15]. A similar evolutionary algorithm was proposed slightly later and independently from us in [66]; this method differs from USPEX in the absence of permutation (i.e. potential problems for binary and more complex compounds), different forms of heredity and mutation, and absence of cell rescaling. We also recently [74,75] developed an approach, enabling deeper insight into the performance of structure prediction simulations (e.g. see below on similarity matrices) and into the energy landscape that is being sampled during the simulation.

Why is the USPEX methodology successful? One of the reasons is that local optimization creates chemically reasonable local environments of the atoms. Another reason is that evolutionary pressure (through selection) forces the population to improve from generation to generation. Yet another reason is the choice of variational operators. In heredity, local arrangements of atoms (spatially coherent pieces of structures) are partly preserved and combined. This respects the predominant short-ranged interactions in crystals and exploits information from the current population. For large systems it may be advantageous to combine slabs of several structures. On the other hand, for systems with very few atoms (or molecules) in the unit cell heredity becomes obsolete (in the limit of 1 atom/unit cell it is completely useless); these cases, however, are trivial for other variation operators and even for local optimization of random structures. As a general note, a successful evolutionary algorithm needs to maintain a balance between the "learning power" and maintaining diversity of the population. Fig. 3 illustrates how, without any prior knowledge, a simulation of boron gradually "learned" about $B_{12}$ icosahedra and arrived at the correct ground-state structure.

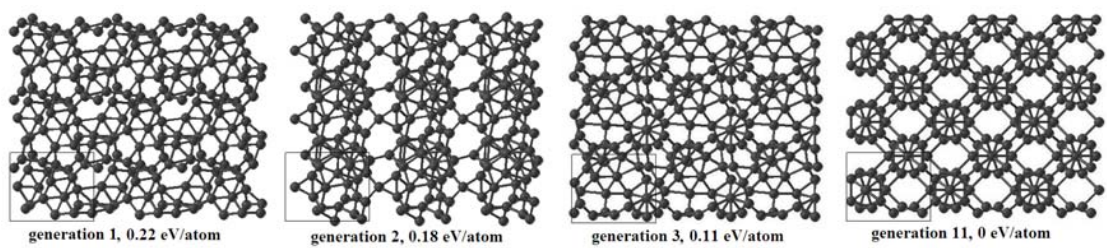

generation 1, 0.22 eV/atom    generation 2, 0.18 eV/atom    generation 3, 0.11 eV/atom    generation 11, 0 eV/atom

**Fig. 3. Illustration of an evolutionary search: 24 atoms of boron in a fixed cell.** The best structure of the first random generation is 0.22 eV/atom above the ground state and is heavily disordered. In the second generation the best structure already contains an incomplete $B_{12}$ icosahedron, the main building block of the ground-state structure. From [17].

Another important reason is that the energy landscapes expected in chemical systems are likely to have an overall "funnel" shape (Fig. 3a), where lowest-energy structures are clustered in the same region of configuration (or order parameter) space. In such cases, evolutionary algorithms are particularly powerful: they "zoom in" on the most promising region of configuration space until the global minimum is found. This "zooming in" is enabled by selection of lower-energy structures as parents for the subsequent generation, and by the form of the variational operators.

Actually, it is possible to test the assumption of an overall benign landscape shape using a recent approach [75] that enables mapping of energy landscapes. If the landscape has one funnel (like in Fig. 4a), there will be a direct correlation between the "distance" of all structures from the ground-state structure (this abstract "distance" measures the degree of structural dissimilarity) and the energy relative to the ground state – indeed, in many real systems (for example, GaAs with 8 atoms/cell – Fig. 4b) such a correlation is found. Even when more than one funnel is present, the number of funnels is usually small (up to three or four). Such situations arise when very different atomic arrangements are energetically competitive, and such systems are particularly challenging as the algorithm may tend to get stuck in one particular funnel. To avoid this, several tools can be used – including dense random or quasirandom sampling (to cover all funnels), tabu lists or special constraint techniques (to deal with each funnel, or a group of funnels, separately).

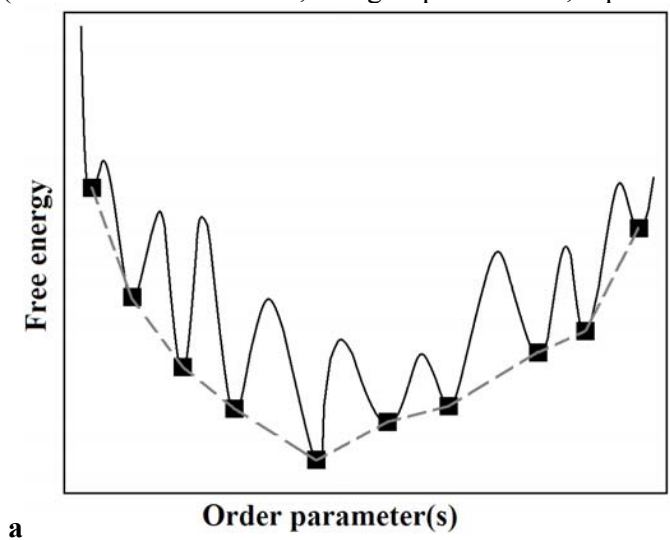

a

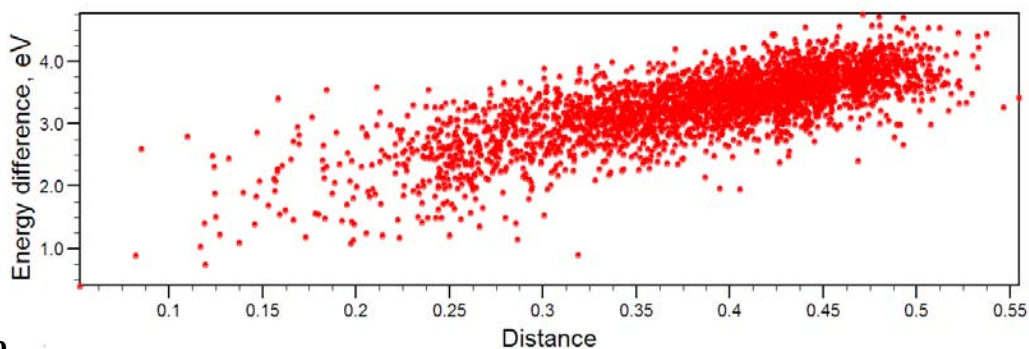

**b**

**Fig. 4. Energy landscapes in chemical systems.** (a) A pedagogical cartoon. The original response surface is very "noisy" (i.e. contains very large energy variations, with high barriers). Local optimization reduces this surface to local minima points (black squares). The reduced response surface (dashed line) is well-behaved and has a simple overall shape. This is one of the reasons why the use of local optimization dramatically improves global optimization [15]. From [76]. (b) Energy-distance correlation for GaAs (8 atoms/cell). Each point is a locally optimized (i.e. relaxed) structure. The correlation proves that the energy landscape has a simple one-funneled topology. From [75].

The energy-distance correlations (Fig. 4b) can be considered as 1D-projections of multidimensional energy landscapes. Projections can, actually, be performed on an arbitrary number of dimensions. Particular visual insight comes from 2D-projections that can be obtained by interpolating and smoothing the 2D-plots presented in [75]. One such depiction of a landscape (for $Au_8Pd_4$ system) is given in Fig. 5.

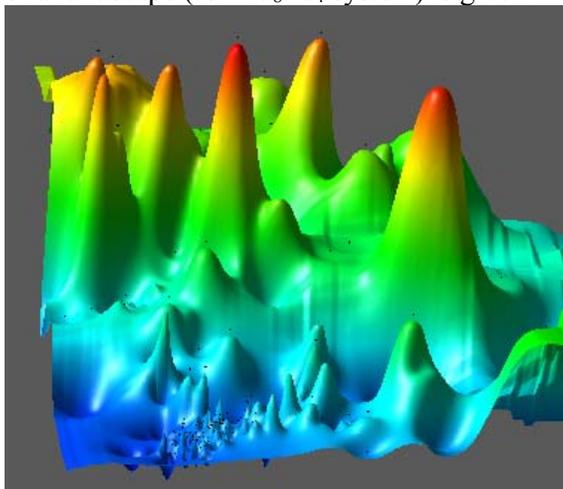

**Fig. 5. 2D-representation of the energy landscape of $Au_8Pd_4$ system using method presented in [75].** The surface has the same meaning as the dashed line in Fig. 4a – it is an interpolation between the points of local minima. Clearly, there is one energy funnel (blue region), which is corresponds to different Au-Pd orderings of the underlying fcc-structure.

The overall landscape shape (Figs. 4, 5) implies that, *en route* to the global minimum some of the low-energy metastable minima can be discovered. This is important, as such phases are often interesting as well. Furthermore, metastable structures found during evolutionary simulations provide a deep insight into the structural chemistry of the studied compound. Thus, evolutionary simulations provide three major results – 1) the ground-state structure; 2) a set of low-energy metastable structures; 3) detailed information on the chemical regime of the compound.

## III. Tests of the algorithm

To measure the strengths and weaknesses of the algorithm, we consider several issues:
1. efficiency of finding the global minimum, in particular relative to a simple well-defined search method, the random sampling,
2. size of systems that can be studied in practice,
3. how fast the diversity decreases along the evolutionary trajectory.

A number of successful tests have been reported in [14,15,32,19,76]. The largest successful test is for a Lennard-Jones crystal with 128 atoms in the (super)cell with variable-cell structure search, which has correctly identified hcp structure as the ground state within 3 generations (each consisting of only 10 structures). For larger Lennard-Jones systems (256 and 512 atoms/cell) we found an energetically very slightly less favourable fcc structure.

The largest test for a chemically complex system is the prediction of the structure of $MgSiO_3$ post-perovskite [34,35] using a relatively large 80-atom supercell (with fixed supercell parameters) and an empirical potential [35] describing interatomic interactions within a partially ionic model. Local optimization and energy calculations were done using the GULP code [31]. Previously [32] we have shown that already in a 40-atom supercell this test is unfeasible using the simple random sampling (with local optimization) [12]: the correct structure was not produced even after $1.2 \times 10^5$ random attempts, but was found with 600-950 local optimizations of structures produced by USPEX. With 80 atoms/cell the problem becomes much more complicated (one expects an exponential increase of complexity with system size), but even in this case we correctly produced the post-perovskite structure in a reasonable number (~3200) of local optimizations – see Fig. 6.

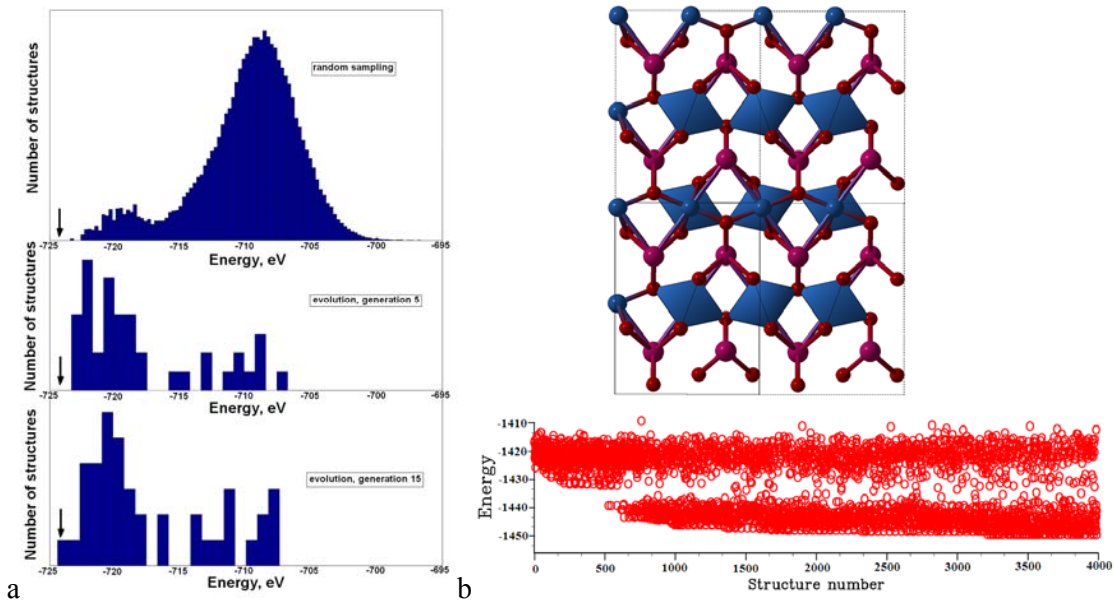

**Fig. 6. Evolutionary prediction of the structure of $MgSiO_3$ post-perovskite using the experimental cell parameters for a) 40-atom [32] and b) 80-atom [19] supercells.** In both cases, each generation consisted of 41 structures. Panel (a) compares densities of states of optimized structures generated randomly (top) and in the evolutionary run. Random sampling did not find the correct structure within $1.2 \times 10^5$ steps, whereas in the evolutionary simulation shown it was found within 15 generations (i.e. 600 local optimizations). Arrows mark the ground-state energy. Panel (b) shows the energies of structures along the evolutionary trajectory for the 80-atom run; the structure of post-perovskite was obtained within ~3200

local optimizations. One can see that the density of low-energy structures increases during the simulation.

Fig. 7 shows variable-cell *ab initio* results for MgSiO$_3$ at the pressure of 120 GPa. Several runs with somewhat different parameters (but within a reasonable range) have been performed and all produced the correct ground-state structure of post-perovskite. The number of local optimizations performed before this structure was found ranged in different runs between 120 and 390; the longest run is shown in Fig. 7.

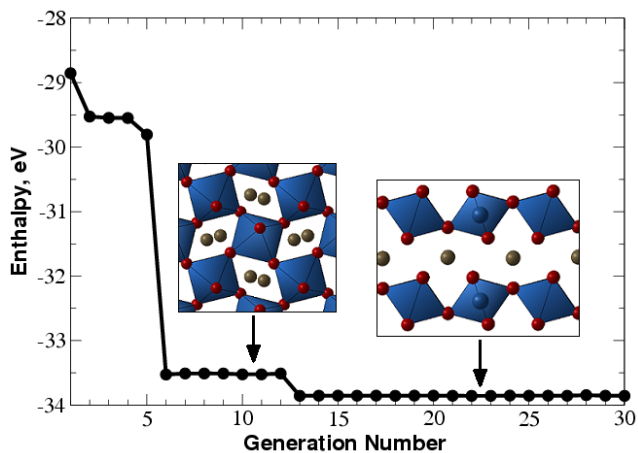

**Fig. 7. Evolutionary structure search for MgSiO$_3$ at 120 GPa**. Evolution of the lowest enthalpy is shown as a function of the generation (insets show the structures of perovskite and post-perovskite phases), from [14].

An example of a very simple test, variable-cell *ab initio* structure search for GaAs with 8 atoms/cell [76], is given in Fig. 8. The ground-state structure for systems of such size can be found even by local optimization of a reasonable number of randomly produced structures. The density of states of relaxed random structures (Fig. 8a), obtained from 3000 random structures, has a characteristic multimodal shape, which seems to be a general feature of energy landscapes. The stable zincblende structure has the abundance of ~0.2%, i.e. finding it with random search would on average take ~ 500 local optimizations. In evolutionary simulations (Fig. 8b) it can be found within 3 generations, or just 30 structure relaxations. Similarity matrices for random (Fig. 8c) and evolutionary (Fig. 8c) searches clearly reveal a strong increase of structure similarity (i.e. decrease of diversity, which can be quantified using the approach [74,75]) along the evolutionary run, after finding the global minimum. Even in this extreme case a significant number of dissimilar structures are produced long after the global minimum is found.

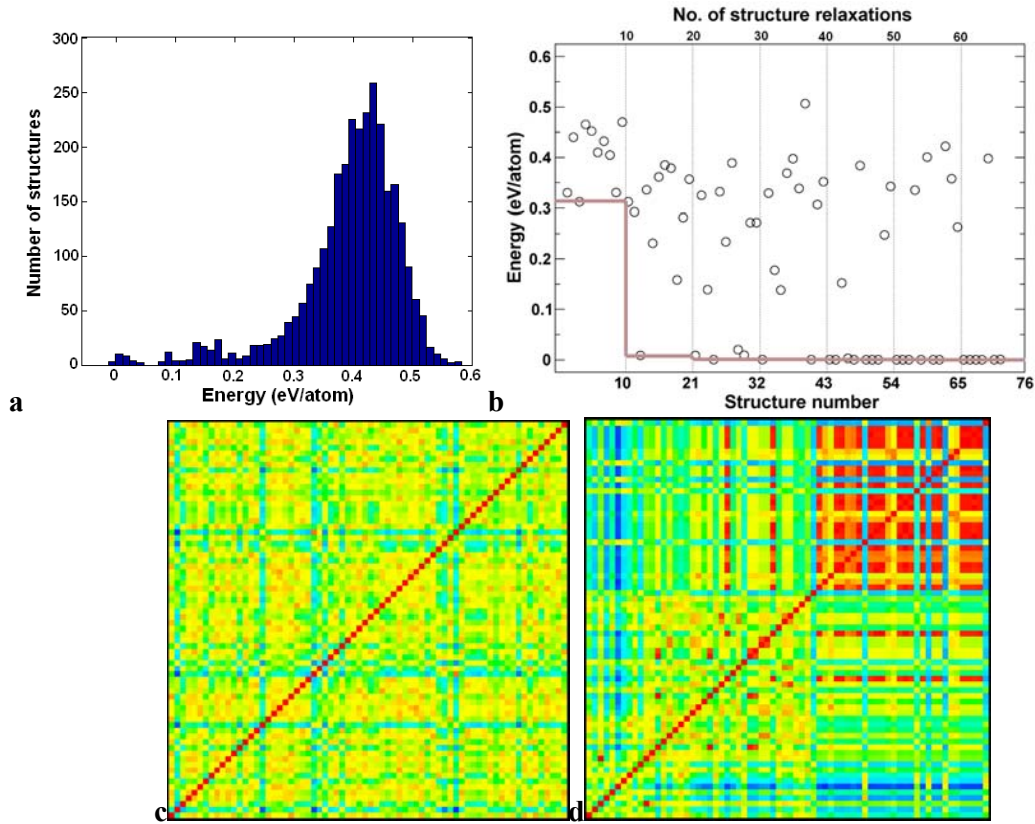

**Fig. 8. Structure prediction for GaAs (8 atoms/cell):** a) energy distribution for relaxed random structures, b) progress of an evolutionary simulation (thin vertical lines show generations of structures, and the grey line shows the lowest energy as a function of generation), c-d) similarity matrices (dimension 70x70) for the random and evolutionary searches, respectively. All energies are relative to the ground-state structure. The evolutionary simulation used a population of 10 structures. Calculations are performed within the GGA [36]. From [76].

$Au_8Pd_4$ (12 atoms/cell) is an unusual system, where a number of different ordered decorations of the fcc structure have competitive energies. The ground state of this system is unknown, but was investigated in several computational studies [71,72,67,76]. Assuming that the ground-state structure should be an ordered variant of the cubic close-packed ("fcc") structure and using the cluster expansion technique with parameters calibrated on a set of *ab initio* energies, Barabash et al. [67] suggested that there are two energetically nearly degenerate structures (Fig. 9c,d). Our calculations found a new ground-state structure (Fig. 9b) that has been overlooked by the previous cluster-expansion study [67] and turned out to be ~0.1 meV/atom lower in energy than the previously known lowest-energy structures (Fig. 9c,d). Examination of all the produced structures shows that most of them are different ordering schemes of the fcc-structure and the energy differences are in most cases very small (Fig. 9a).

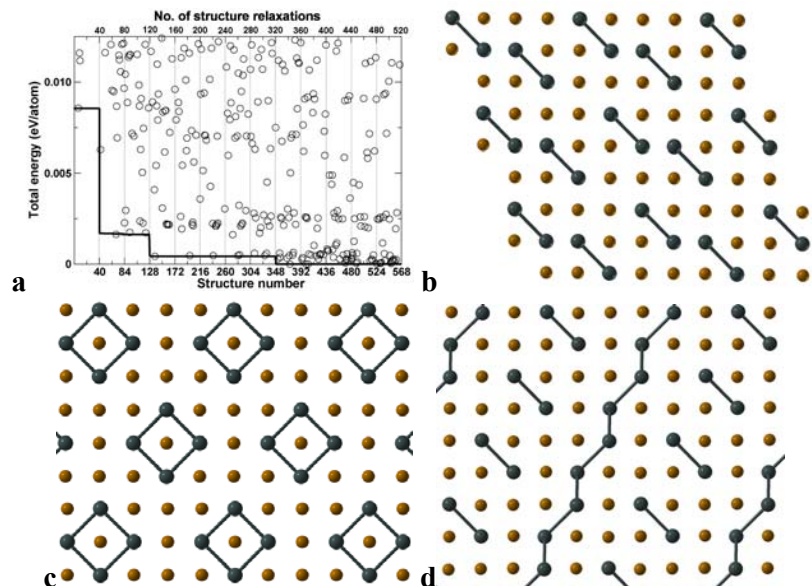

**Fig. 9. Evolutionary structure search for $Au_8Pd_4$**. a) evolution of the total energies (only the lowest-energy part is shown for clarity), b) the lowest-energy structure found in our evolutionary simulation, c-d) – the lowest-energy structures found by cluster expansion in Ref. [67] (structures No. 4905 and No. 4557, respectively, in the search catalogue of Ref. [67]). Energies are given relative to the ground state.

Periodic boundary conditions suppress decomposition, but when a compound is extremely unstable against decomposition, phase separation can be observed in USPEX simulations. Actually, this happens rather frequently in explorations of hypothetical compositions. A clear example is given by the Cu-C system, which does not have any stable compounds. The tendency to unmixing in this system is very strong and even simulations on small cells show clear separation into layers of fcc-structured Cu and layers of graphite (Fig. 10). When the tendency to unmixing is not so large, simulations on small unit cells may find metastable "mixed" structures. Such structures have the lowest thermodynamic potential only at the given number of atoms in the unit cell; increasing the cell size would lead to phase separation. In the Cu-C system, phase separation is evident already at very small system sizes (Fig. 10).

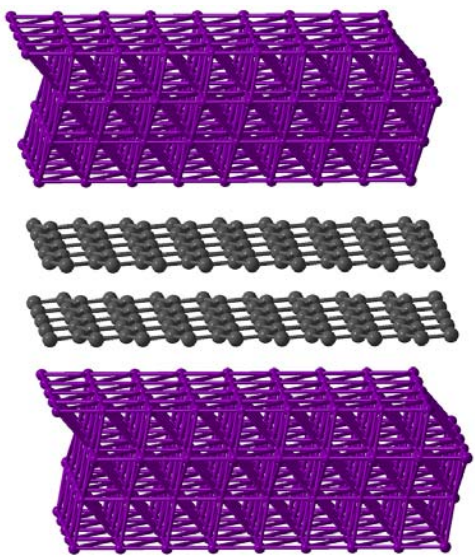

**Fig. 10. Lowest-energy structure of Cu$_2$C with 12 atoms/cell at 1 atm.**

### IV. Some applications of the method

In this section we will review some new insight that has been obtained using our method (see also [14]). All structure predictions described here were performed within the generalized gradient approximation (GGA [36]) and the PAW method [37,38], using VASP code [29] for local optimization and total energy calculations. The predicted structures correspond to the global minimum of the approximate free energy surface. For systems where the chosen level of approximation (GGA in cases considered below) is adequate, this corresponds to the experimentally observed structure. Where this is not the case, results of global optimization are invaluable for appraising the accuracy of the approximations.

**CaCO$_3$ polymorphs.** High-pressure behaviour of carbonates is very important for the global geochemical carbon cycle, as high-pressure carbonates of Mg and Ca are expected to contain most of the Earth's carbon [40]. For CaCO$_3$, there is a well-known transition from calcite to aragonite at ~2 GPa, followed by a transition to a post-aragonite phase at ~40 GPa [42], the structure of which was solved [13] using USPEX, and the predicted structure matched the experimental X-ray diffraction pattern well. Furthermore, we have predicted [13] that above 137 GPa a new phase, with space group $C222_1$ and containing chains of carbonate tetrahedra, becomes stable. Recently this prediction was verified by experiments [43] at pressures above 130 GPa. We note that both post-aragonite and the $C222_1$ structure (Fig. 11) belong to new structure types and could not have been found by analogy with any known structures.

The presence of tetrahedral carbonate-ions at very high pressures invites an analogy with silicates, but the analogy is limited. In silicates, the intertetrahedral angle Si-O-Si is extremely flexible [45], which is one of the reasons for the enormous diversity of silicate structure types. Fig. 12 shows the variation of the energy as a function of the Si-O-Si angle in the model H$_6$Si$_2$O$_7$ molecule – method borrowed from [45]. One can see only a shallow minimum at $\angle(Si-O-Si)$=135º, but a deep minimum at $\angle(C-O-C)$=124º with steep energy variations for H$_6$C$_2$O$_6$ (Fig. 12). This suggests a much more limited structural variety of metacarbonates, compared to silicates. In both CaCO$_3$ and CO$_2$ the $\angle(C-O-C)$ angles are close to 124º in a wide pressure range.

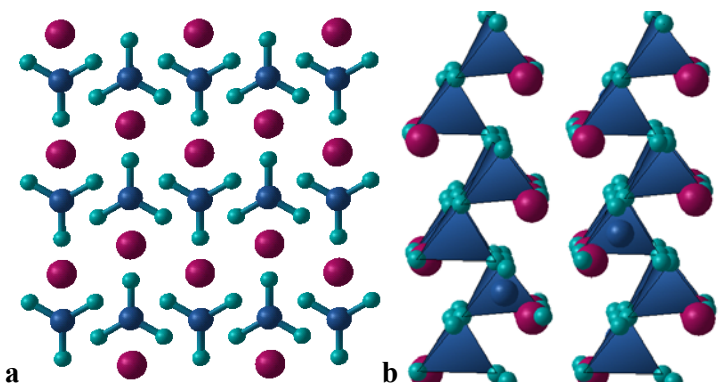
**Fig. 11. CaCO$_3$ at high pressure**. a) structure of post-aragonite phase, b) $C222_1$ phase.

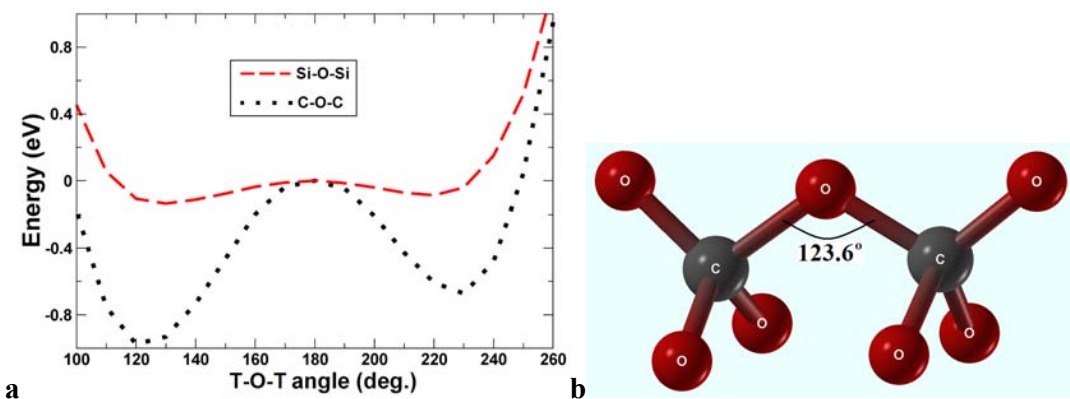
**Fig. 12. Energy variation as a function of the T-O-T angle** (red dashed line – T = Si, black dotted line – T = C). Calculations were performed on H$_6$T$_2$O$_7$ molecules; at each angle all T-O distances and O-T-O valence angles were optimized. Optimum angle C-O-C=124º, Si-O-Si=135º. These calculations were performed with SIESTA code [30] using the GGA functional [36], norm-conserving pseudopotentials and a double-ζ basis set with a single polarization function for each atom.

**Polymeric phase of CO$_2$ [47].** High-pressure behaviour of CO$_2$ is still controversial [48]. It is known that above ~20 GPa a non-molecular phase (called phase V [49]) with tetrahedrally coordinated carbon atoms becomes stable, but its structure is still under debate: in the first experimental study [49] a trydimite structure was proposed, but later theoretical works found it to be unstable (even not metastable) and much less favourable than the β-cristobalite structure [50,51]. At the same time, it was not possible to rule out that there may be even more stable structures. We have performed [47,76] evolutionary structure searches at 50 GPa, 100 GPa and 150 GPa for systems with 6, 9, 12, 18 and 24 atoms/cell. At all these pressures we confirmed stability of the β-cristobalite structure (Figs. 13 and 14), thus suggesting an experimental re-investigation of phase V of carbon dioxide. CO$_2$-V is stable against decomposition into diamond and oxygen (the enthalpy of decomposition is very large and increases from 3.3 eV to 3.8 eV between 50 GPa and 200 GPa).

At lower pressures, between 8.9 GPa and 18.9 GPa, the $P\frac{4_2}{m}nm$ phase (see [64] for details) is stable, and at even lower pressures (0-8.9 GPa) the $Pa3$ structure is stable (Fig. 13). The

$Pa3$-$P\frac{4_2}{m}nm$ transition pressure calculated here (8.9 GPa) is consistent with experiment and previous calculation [64].

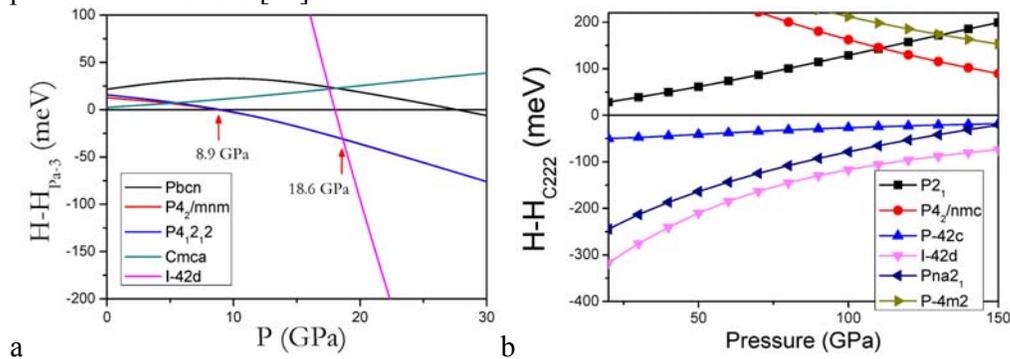

**Fig. 13. Enthalpies of candidate forms of $CO_2$:** a) in the low-pressure region, relative to the molecular $Pa3$ structure, b) in the high-pressure region, relative to the non-molecular $C222$ structure. From [47].

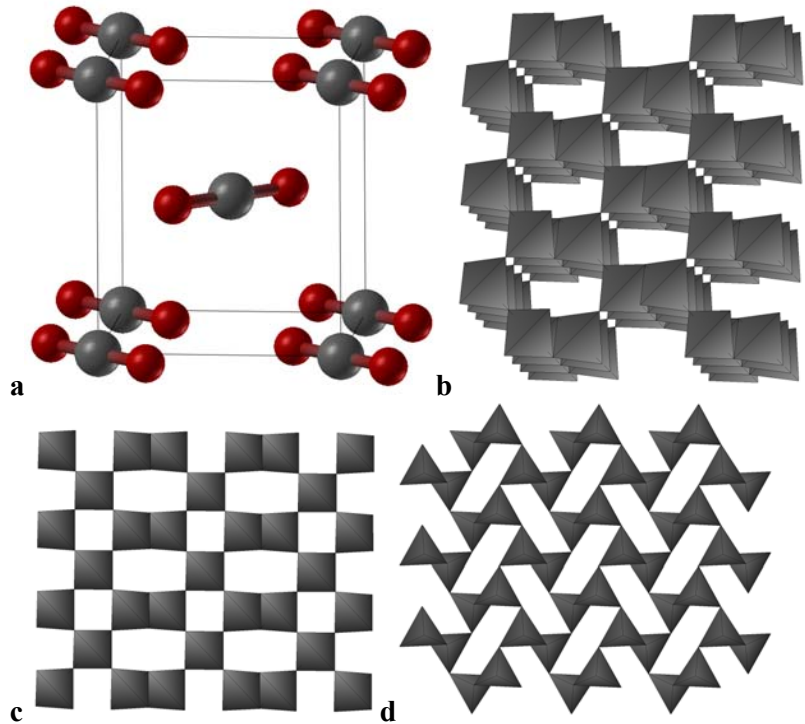

**Fig. 14. $CO_2$ structures:** a) molecular $P\frac{4_2}{m}nm$ structure, stable at lower pressures than $CO_2$-V, b) polymeric β-cristobalite-type form of $CO_2$, suggested to be the structure of phase V and showing carbonate tetrahedra. Structural parameters at 100 GPa: space group $I\bar{4}2d$, $a=b=3.2906$ Å, $c=6.0349$ Å, C(0.5; 0; 0.25), O(0.2739; 0.25; 0.125), c) polymeric $C222$ structure, d) metastable polymeric $Pna2_1$ structure. From [47].

**Semiconducting and metallic phases of solid oxygen: unusual molecular associations.**
The red ε-phase of oxygen, stable in the pressure range 8-96 GPa, was discovered in 1979

[52], but its structure was solved only in 2006 [54,55]. The metallic (superconducting at very low temperatures [57]) ζ-phase, stable above 96 GPa, was discovered in 1995 [53], and its structure remained controversial for a long time. Neutron diffraction showed [58] that already in the ε-phase (at 8 GPa) there is no long-range magnetic order and likely even no local moments. The disappearance of magnetism is a consequence of increasing overlap of molecular orbitals with increasing pressure. Ultimately, orbital overlap leads to metallization. To understand high-pressure chemistry of oxygen, we performed extensive structure searches at pressures between 25 GPa and 500 GPa [14,16], taking into account only non-magnetic solutions.

At 25 GPa we found two particularly interesting structures – one consisting of zigzag chains of $O_2$ molecules (*Cmcm* structure of Ref. [60,14], see Fig. 15b) and one with more complex chains of molecules (see Fig. 15c). These have strong similarities with the experimentally observed structure (Ref. [54,55], see Fig. 15a) consisting of $O_8$ clusters: all of these structures are molecular, and in all of them each molecule is connected with two other molecules, at distances of ~2.1-2.2 Å (the intermolecular distance is ~1.2 Å). The *Cmcm* structure, first suggested in [60], is the true GGA ground state, but it differs from experiment; as Fig. 16a shows, its enthalpy is ~10 meV/atom lower than for the experimentally found structure (Fig. 15a). Metastability of the experimentally studied structure cannot yet be ruled out, but it seems likely that this discrepancy is due to deficiencies of the GGA. The $(O_2)_4$ clusters are held together by weak intermolecular covalent bonds: each $O_2$ molecule has two unpaired electrons occupying two molecular π*-orbitals, and sharing these electrons with neighbouring molecules creates two intermolecular bonds per molecule and a non-magnetic ground state [16,62]. It is well known that DFT-GGA does not perform well for stretched covalent bonds, the root of the problem being in the locality of the exchange-correlation hole in DFT-GGA, whereas the true exchange-correlation hole in such cases is highly delocalized. At high pressure, intermolecular distances decrease, intermolecular bonds become more similar to normal covalent bonds and the true exchange-correlation hole becomes more localized. Therefore, we can apply the GGA with greater confidence for the prediction of the structure of the metallic ζ-phase.

For the ζ-phase, evolutionary simulations at 130 GPa and 250 GPa uncovered two interesting structures with *C*2/*m* and *C*2/*c* space groups [16]. These have very similar enthalpies (Fig. 15a); the *C*2/*m* structure is slightly lower in enthalpy and matches experimental X-ray diffraction and Raman spectroscopy data very well, better than the *C*2/*c* structure [16]. Recently (Weck & Loubeyre, pers. comm.) our predicted *C*2/*m* structure of ζ-oxygen was experimentally confirmed. Both structures contain well-defined $O_2$ molecules; our simulations show that oxygen remains a molecular solid at least up to 500 GPa. Phonon dispersion curves of the *C*2/*m* structure (Fig. 15b-d) contain clearly separated molecular vibrons and show that the structure is dynamically stable (Fig. 16), except at 110 GPa, where we see tiny imaginary frequencies in the Γ-V direction, close to the Brillouin zone centre. Such soft modes may result in small-amplitude long-wavelength modulations of the structure at very low temperatures.

The ε-ξ transition is isosymmetric, which implies that it is first-order at low temperatures but can become fully continuous above some critical temperature [61]. Given the small volume discontinuity upon transition and small hysteresis (one can obtain the *C*2/*m* structure of the ξ-phase by simple overcompression of the ε-$O_8$ structure, ~5 GPa above the thermodynamic transition pressure), one can expect this critical temperature to be rather low. We note that within the GGA the ε-ξ transition is predicted to occur at 45 GPa (Fig. 16), much lower than the experimental transition pressure (96 GPa). This has two explanations – (i) as the GGA is expected to perform better for the metallic ζ-phase than for the semiconducting ε-$O_8$ phase, the enthalpy differences are expected to suffer from non-cancelling errors, (ii) since the ε-ξ transition is not only structural, but also electronic (insulator-metal transition), one might expect metallization at lower pressures than in experiment. Typically, density functional calculations overestabilize metallic states relative to insulating ones, and this is exactly what we observe in oxygen. Very recent single crystal experiments [107] have confirmed our predicted $C2/m$ structure for ζ-oxygen.

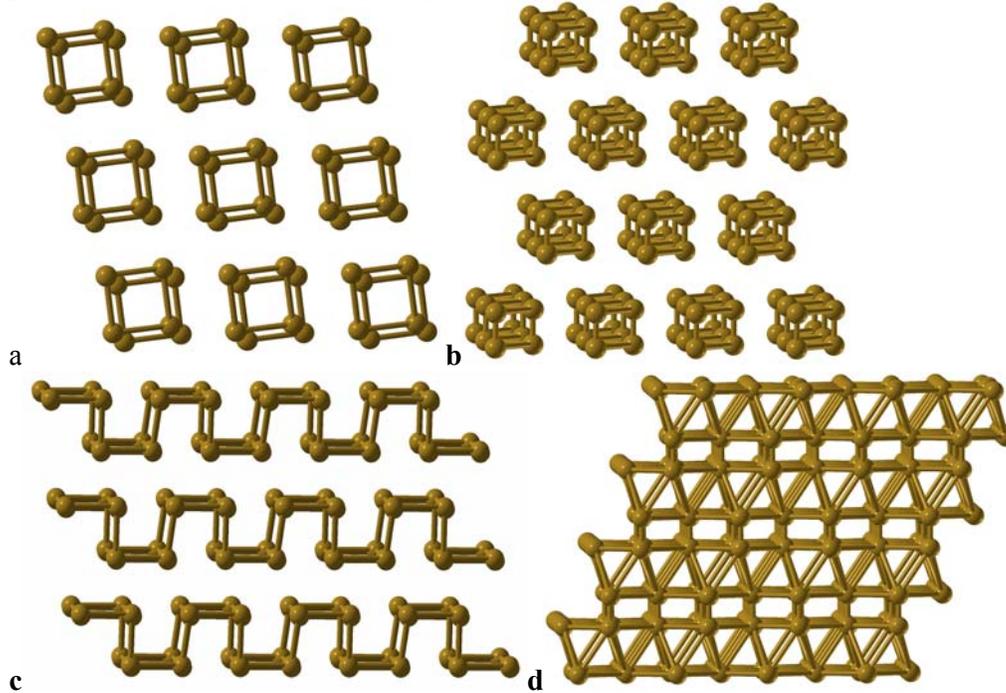

**Fig. 15. High-pressure structures of oxygen:** a) experimentally found ε-$O_8$ structure at 17.5 GPa [54], b) $Cmcm$ chain structure [60,14], c) metastable $P\bar{1}$ chain structure at 25 GPa [14], d) $C2/m$ structure of the ζ-phase at 130 GPa [16]. Contacts up to to 2.2 Å are shown as bonds. From [76].

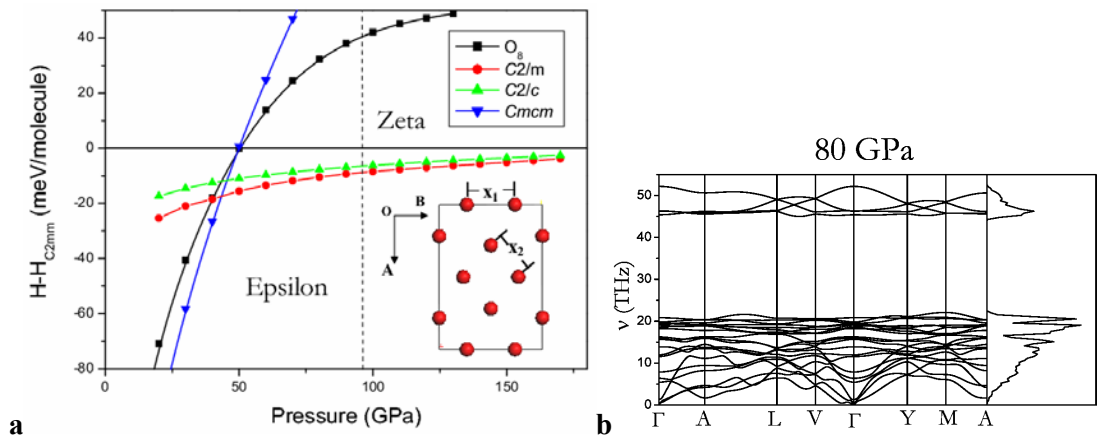

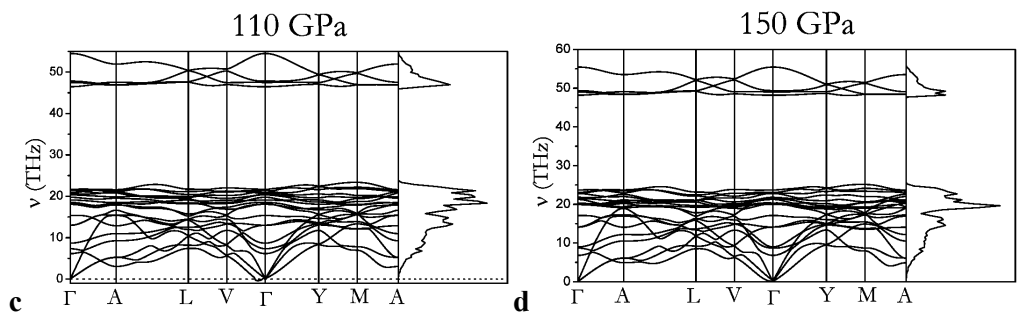

**Fig. 16. High-pressure phases of oxygen:** a) enthalpies (relative to the *C2mm* structure of Ref. [59]) of several possible structures as a function of pressure (from [16]) of several possible structures, b) phonon dispersion curves and densities of states of the *C2/m* ζ-phase at three pressures. From [76].

**Reactivity of noble gases: are Xe-C compounds possible at high pressure?** Inducing major changes in the electronic structure of atoms, high pressure may also change their reactivity. For instance, noble (i.e. largely unreactive) metal platinum under pressure easily forms carbide PtC [77,78] and dinitride $PtN_2$ [79]. One should not confuse chemical reactivity with propensity to phase transitions: recently [80] it was concluded that gold loses its "nobility" at 240 GPa, when it undergoes an fcc-hcp structural transition. Structural transitions and reactivity are not linked, however: in spite of becoming reactive, Pt does not change its fcc structure, and Cu (not a noble metal by any standards) is only known in one crystalline phase with the fcc structure.

An interesting question is whether noble gases become reactive. Indeed, it was observed that a few percent Xe can be incorporated in quartz ($SiO_2$) at elevated pressures and high temperatures [39]. A possibility has been suggested [40] that stable Xe-C compounds may be stable at high pressure; indeed, carbon and xenon have similar valence orbital energies (*cf.* ionization potentials of 12.13 eV and 11.26 eV for Xe and C, respectively) and one expects that pressure would make Xe more reactive [39]. We did simulations at 200 GPa, i.e. above the metallization pressure of Xe (132 GPa [70]), when its closed electronic shells are strongly perturbed. These calculations were done within the GGA [36] and on cells containing up to 14 atoms/cell. At this pressure all Xe carbides are extremely unstable (Fig. 17) and their structures show clear separation into close-packed Xe layers (i.e. fragments of the elemental Xe structure) and 3,4-connected carbon layers (intermediate between graphite and diamond). The only exception is the 3D-clathrate structure of $XeC_8$. The observed layering is consistent with the instability to decomposition. Although Xe carbides are unstable at 200 GPa, already at that pressure we observe considerable bonding Xe-C interactions and the effect of Xe on the carbon sublattice is far beyond simple mechanistic size factor – the carbon layers adopt unusual and exciting configurations that may be prepared in the laboratory under certain conditions.

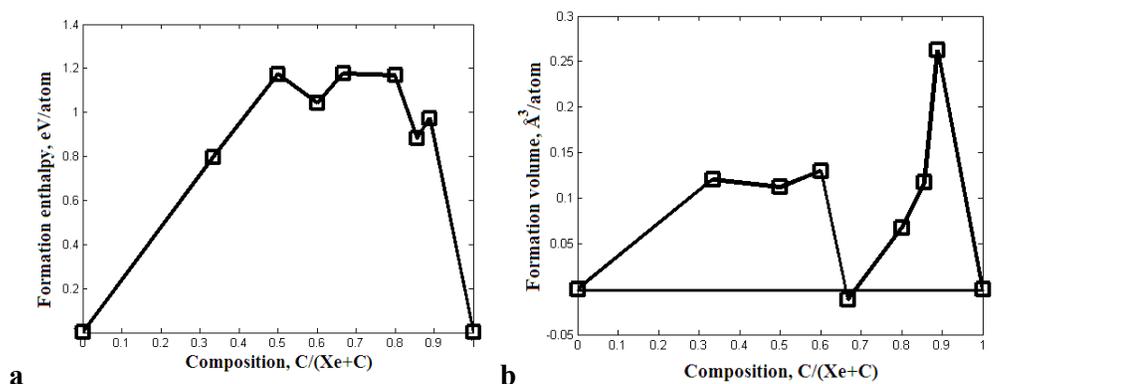

**Fig. 17. Predicted enthalpy (a) and volume (b) of formation of Xe-C compounds at 200 GPa.** The compounds shown are Xe (hcp), $Xe_2C$, XeC, $Xe_2C_3$, $XeC_2$, $XeC_4$, $XeC_6$, $XeC_8$, C(diamond). Note that $XeC_2$ has a small negative volume of formation and might become stable at much higher pressures. From [76].

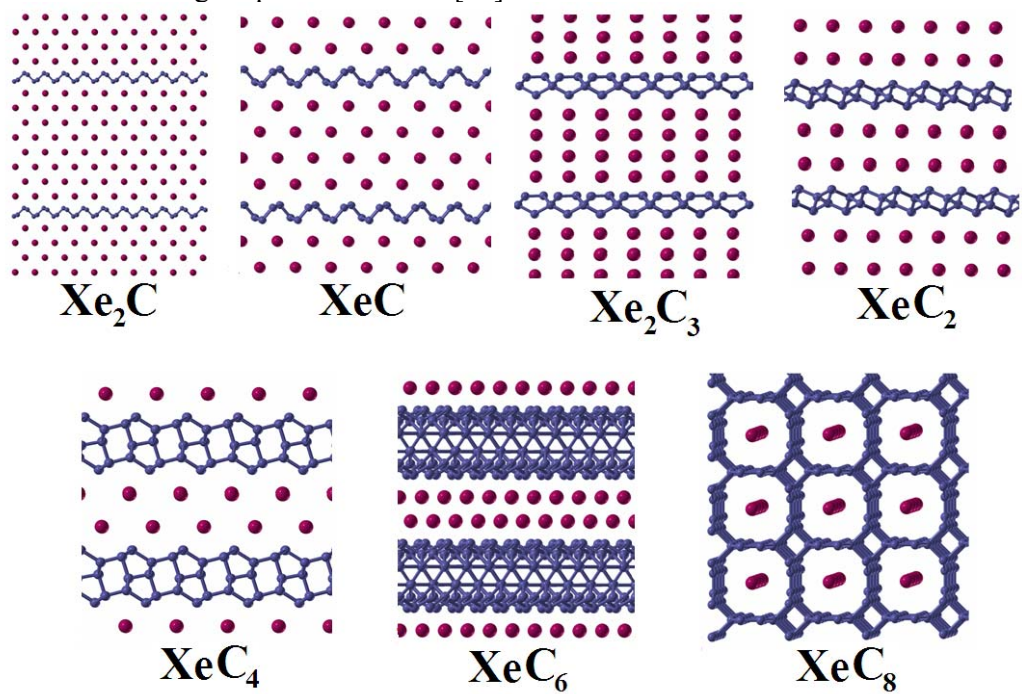

**Fig. 18. Predicted structures of $Xe_2C$, XeC, $Xe_2C_3$, $XeC_2$, $XeC_4$, $XeC_6$, $XeC_8$ at 200 GPa.** From [76].

**Boron: novel phase with a partially ionic character.** Boron is perhaps the most enigmatic element: at least 16 phases were reported in the literature, but most are believed or suspected to be compounds (rather than forms of the pure element), and until recently the phase diagram was unknown. Following experimental findings of J. Chen and V.L. Solozhenko (both arrived independently at the same conclusions in 2004) of a new phase at pressures above 10 GPa and temperatures of 1800-2400 K, whose structure could not be determined from experimental data alone. We found the structure using USPEX. We named this phase γ-$B_{28}$ (because it contains 28 atoms/cell). Its structure has space group *Pnnm* and is comprised of icosahedral $B_{12}$ clusters and $B_2$ pairs in a NaCl-type arrangement. This phase is stable between 19 and 89 GPa, and exhibits sizable charge transfer from $B_2$ pairs to $B_{12}$ clusters, quite unexpected for a pure element. Details are given in Ref. [17] and Figs. 19,20.

Fig. 21 shows a comparison of theoretical and experimental X-ray powder diffraction profiles.

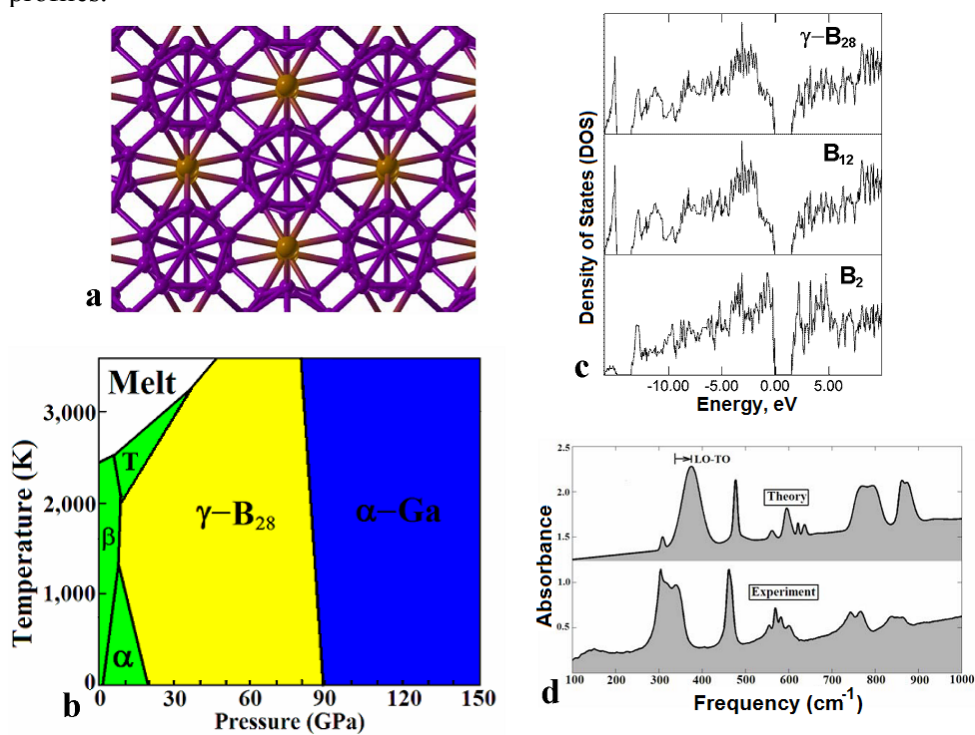

**Fig. 19. Boron:** (a) structure of γ-$B_{28}$ ($B_{12}$ icosahedra are shown in blue, $B_2$ pairs – in orange). (b) phase diagram of boron, showing a wide stability field of γ-$B_{28}$. (c) electronic DOS and its projections onto $B_{12}$ and $B_2$ units (all DOSs are normalized per atom), (d) comparison of theoretical and experimental IR spectra. IR spectra indicate the presence of non-zero Born charges on atoms. From [17].

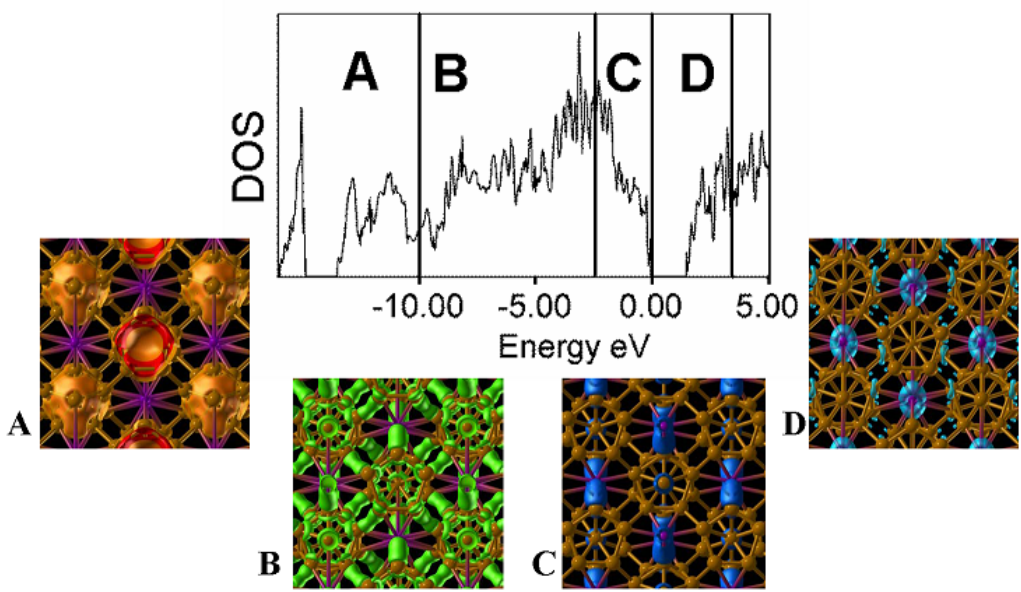

**Fig. 20. γ-$B_{28}$: total electronic DOS and energy-decomposed electron densities.** Lowest-energy valence electrons are dominated by the $B_{12}$ icosahedra, while top of the valence band and bottom of the conduction band (i.e. holes) are localized on the $B_2$

pairs. This is consistent with atom-projected DOSs (Fig. 19 c) and the idea of charge transfer $B_2 \rightarrow B_{12}$.

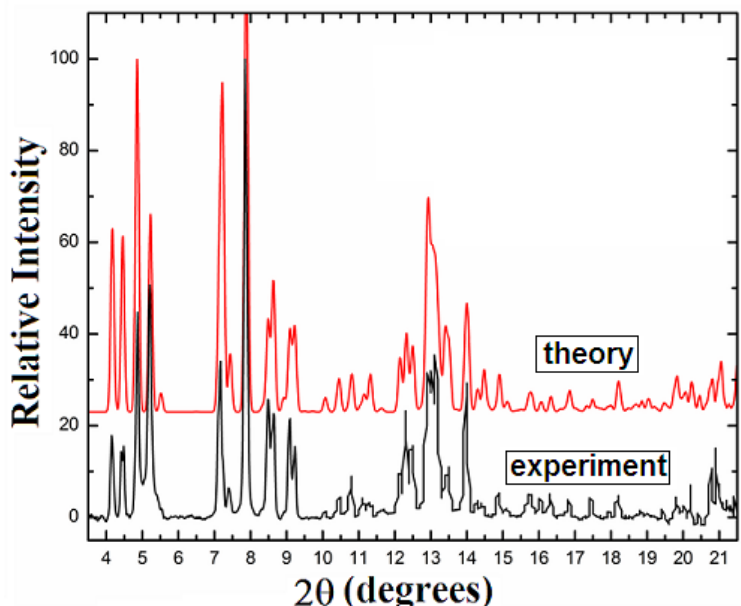

**Fig. 21. Comparison of theoretical and experimental X-ray powder diffraction profiles of γ-$B_{28}$.** X-ray wavelength λ = 0.31851 Å. From [17].

γ-$B_{28}$ can be represented as a "boron boride" $(B_2)^{\delta+}(B_{12})^{\delta-}$; although the exact value of the charge transfer δ depends on the definition of an atomic charge, for all definitions that we used the qualitative picture is the same. Perhaps the most reliable definition of a charge, due to Bader [81], gives δ~0.5 [17]. We have recently found that (based on the similarity of synthesis conditions and many diffraction peaks) the same high-pressure boron phase may have been observed by Wentorf in 1965 [82], though Wentorf's material was generally not believed to be pure boron (due to the sensitivity of boron to impurities and lack of chemical analysis or structure determination in [82]) and its diffraction pattern was deleted from Powder Diffraction File database. γ-$B_{28}$ is structurally related to several compounds – for instance, $B_6P$ [83] or $B_{13}C_2$ [84]), where the two sublattices are occupied by different chemical species (instead of interstitial $B_2$ pairs there are P atoms or C-B-C groups, respectively). Significant charge transfer can be found in other elemental solids, and observations of dielectric dispersion [85], equivalent to LO-TO splitting, suggest it for β-$B_{106}$. The nature of the effect is possibly similar to γ-$B_{28}$. Detailed microscopic understanding of charge transfer in β-$B_{106}$ would require detailed knowledge of its structure, and reliable structural models of β-$B_{106}$ finally begin to emerge from computational studies [86,87]. It is worth mentioning that γ-$B_{28}$ is a superhard phase, with a measured Vickers hardness of 50 GPa [88], which puts it among half a dozen hardest materials known to date.

Nearly two years after the original work [17,88], Zarechnaya's group confirmed the synthesis, crystal structure, and superhardness of this phase [104,102]. However, the reported contamination of their samples by Pt (from the capsules in which they performed synthesis) decreases the value of their results. Their measured hardness of 58 GPa [104] is almost certainly an overestimate. There are a number of other problems with their study – e.g. they estimate the density difference between γ-$B_{28}$ and other boron phases as 1%, whereas in reality it is much greater (e.g., γ-$B_{28}$ is 8.3% denser than β-$B_{106}$), and their interpretation of chemical bonding as fully covalent is subject to criticism. Mixed ionic-covalent bonds, as in γ-$B_{28}$, are well known in chemistry (e.g. Si-O or C-O bonds). Perhaps unaware of this, Zarechnaya et al. [104], having found evidence for a covalent component, incorrectly concluded that there can be no ionic component. This is equivalent to saying that, since there is an accumulation of charge density on Si-O bonds, the bond is fully covalent – which, of course, is not the case. A summary of work on γ-$B_{28}$ is given in Table 1.

**Table 1. Summary of work on γ-$B_{28}$ phase.**

| Reference | Main findings | Submission date |
|---|---|---|
| Wentorf [82] | Synthesis, density measurement, qualitative X-ray diffraction, electrical conductivity change across β-γ transition. No chemical analysis and no structure determination. | 04.10.1964 |
| Oganov et al. [17] | Synthesis and proof of chemical purity, structure determination, demonstration of partial ionicity, phase diagram. | 27.01.2007 |
| Solozhenko et al. [88] | Hardness measurement. | 03.10.2009 |
| Zarechnaya et al. [102] | Confirmation of synthesis and structure. Identified γ-$B_{28}$ with Wentorf's phase. | 03.11.2009 |
| Le Godec et al. [103] | Accurate measurement of the equation of state at 300 K. | 08.01.2009 |
| Zarechnaya et al. [104,102] | Re-confirmation of structure using single crystals. Measurements of the band gap and electrical conductivity. Inaccurate measurements of the equation of state and hardness at 300 K. Incorrect interpretations of chemical bonding density differences in boron phases. | 16.01.2009 |
| Jiang et al. [105] | Simulation of structure deformation. | 12.03.2009 |
| Rulis et al. [106] | Simulation of electronic spectra, supporting charge transfer picture. | 06.04.2009 |

**Sodium: a metal that goes transparent under pressure.** A sequence of recent discoveries demonstrated that sodium, a simple s-element at normal conditions, behaves in highly non-trivial ways under pressure. The discovery of an incommensurate host-guest structure [89], followed by finding of several complex structures [90] in the range of pressures corresponding to the minimum of the melting curve [91], and the very existence of that extremely deep minimum in the melting curve at about 110 GPa – all this evidence points to some unusual changes in the physics of sodium. Later it was shown also that the incommensurate host-guest structure is a 1D-metal [92], where conductivity is primarily due to chains formed by the guest sublattice. Yet another unusual phenomenon was predicted using USPEX and later (but within the same paper [93]) verified experimentally: on further compression sodium becomes a wide-gap insulator! This happens at ~190 GPa, and Fig. 22 shows the crystal structure of the insulating "hP4" phase, its enthalpy relative to other structures, and the electronic structure. The structure can be described as a double hexagonal close-packed (dhcp) structure, squeezed by a factor of >2 along the *c*-axis, as a result of which sodium atoms have 6-fold coordination. There are 2 inequivalent Na positions: Na1 and Na2, which have the octahedral and trigonal-prismatic coordination, and the hP4 structure can be described as the elemental analog of the NiAs structure type (the same way as diamond is the elemental analog of the zincblende structure type). Calculations suggest that sodium is no longer an s-element; instead, the outermost valence electron has significant s-, p- and d-characters (Fig. 22c). In other words, sodium can be considered as a transition metal, because of its significant d-character.

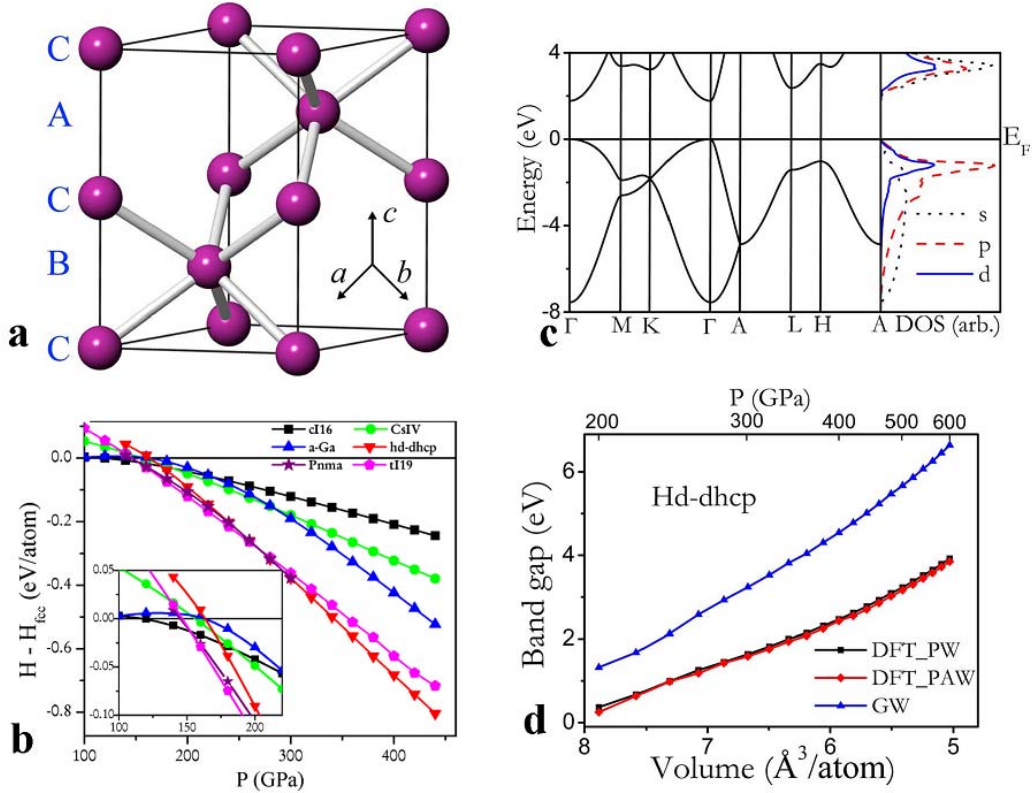

**Fig. 22. Summary on the hP4 phase of sodium:** a) its crystal structure, b) enthalpies of competing high-pressure phases (relative to the fcc structure), c) band structure, d) pressure dependence of the band gap, indicating rapid increase of the band gap on compression. From [93].

The band gap is direct, and increases with pressure. At 200 GPa the bandgap calculated with the GW approximation (known to give rather accurate results) is 1.3 eV, and increases to 6.5 eV at 600 GPa. These predictions implied that above 200 GPa sodium will be red and transparent, and at ~300 GPa it will become colorless and transparent (like wide-gap insulators). This has indeed been confirmed in experiments of M.I. Eremets [93], see Fig. 23. The insulating behavior is explained by the extreme localization of the valence electrons in the interstices of the structure, i.e. the "empty" space (Fig. 24). These areas of localization are characterized by surprisingly high values of the electron localization function (nearly 1.0) and maxima of the total electron density. The number of such maxima is half the number of sodium atoms, and therefore in a simple model we can consider Na atoms as completely ionized ($Na^+$), and interstitial maxima as containing one electron pair. The hP4 structure can also be described as a $Ni_2In$-type structure, where Na atoms occupy positions of Ni atoms, and interstitial electron pairs in hP4-Na sit on the same positions as In atoms in $Ni_2In$. At first counterintuitively, the degree of localization of the interstitial electron pairs increases with pressure, explaining the increase of the band gap (Fig. 22d). hP4-Na can be described as an electride, i.e. an ionic "compound" formed by ionic cores and localized interstitial electron pairs. The very fact that sodium, one of the best and simplest metals, under pressure becomes a transparent insulator with localized valence electrons, is remarkable and forces one to reconsider classical ideas of chemistry.

Interstitial charge localization can be described in terms of (s)-p-d orbital hybridizations, and its origins are in the exclusionary effect of the ionic cores on valence electrons: valence electrons, feeling repulsion from the core electrons, are forced into the interstitial regions at pressures where atomic cores begin to overlap [94].

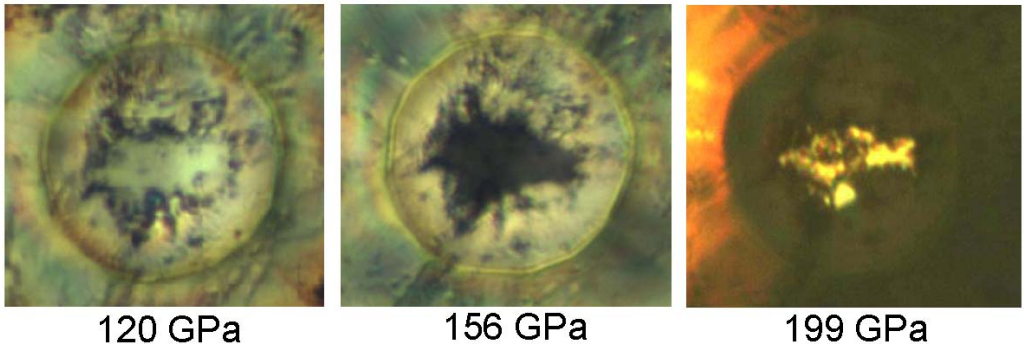

**Fig. 23. Photographs of sodium samples under pressure.** At 120 GPa the sample is metallic and highly reflective, at 156 GPa the reflectivity is very low, and at 199 GPa the sample is transparent. From [93].

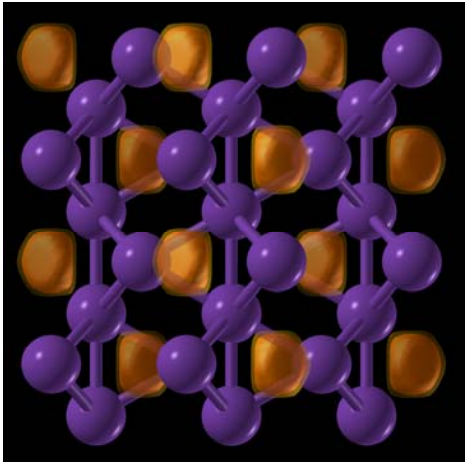

**Fig. 24. Crystal structure and electron localization function (orange isosurfaces - lowest contour 0.82) of the hP4 phase of sodium at 300 GPa.** Interstitial electron localization is clearly seen.

## V. Conclusions

Evolutionary algorithms, based on physically motivated forms of variation operators and local optimization, are a powerful tool enabling reliable and efficient prediction of stable crystal structures. This method has a wide field of applications in computational materials design (where experiments are time-consuming and expensive) and in studies of matter at extreme conditions (where experiments are very difficult or sometimes beyond the limits of feasibility).

One of the current limitations is the accuracy of today's *ab initio* simulations; this is particularly critical for strongly correlated and for van der Waals systems. Note, however, that the method itself does not make any assumptions about the way energies are calculated and can be used in conjunction with any method that is able to provide total energies. Most of practical calculations are done at T=0 K, but temperature can be included as long as the free energy can be calculated efficiently. Difficult cases are aperiodic and disordered systems (for which only the lowest-energy periodic approximants and ordered structures can be predicted at this moment).

We are suggesting USPEX as the method of choice for crystal structure prediction of systems with up to ~50 atoms/cell, where no information (or just the lattice parameters) is available. Above 50-100 atoms/cell runs become expensive (although still feasible), eventually necessitating the use of other ideas within USPEX or another approach, due to the 'curse of dimensionality'. There is hope of enabling structure prediction for very large (> 200 atoms/cell) systems. The extension of the method to molecular systems (i.e. handling whole molecules, rather than individual atoms) is already available. The first successful step has been made [95] in adapting USPEX to clusters. Similar extensions, relatively straightforward, still need to be done for surfaces and interfaces. One major unsolved problem is the simultaneous prediction of all stable stoichiometries and structures (in a given range of compositions). A pioneering study [73] succeeded in predicting stable stoichiometries of alloys within a given structure type, but simultaneous prediction of structure and stoichiometry remains an unsolved problem.

USPEX has been applied to many important problems. Apart from the applications described above, several noteworthy results have been published by us recently. These include the high-pressure post-magnesite phases of $MgCO_3$ [47], polymeric phases of nitrogen [96], superconducting phases of silane ($SiH_4$) [97] and germane ($GeH_4$) [98], the latter predicted to have a remarkably high $T_C$=64 K [98]. Its ability to predict not only the ground states, but also low-energy metastable structures has led to the finding [14] of an interesting metastable structure of carbon, which has recently been shown [100] to match the observed properties of the so-called "superhard graphite", a material scratching on diamond and formed by metastable room-temperature compression of graphite beyond 15 GPa [101]. One expects many more applications to follow, both in high-pressure research and in materials design.

**Acknowledgements**
ARO thanks R. Hoffmann, W. Grochala, R.J. Hemley and R.M. Hazen for exciting discussions. ARO also gratefully acknowledges financial support from the Research Foundation of Stony Brook University and from Intel Corporation. YM's work is supported by the China 973 Program under Grant No. 2005CB724400, the National Natural Science Foundation of China under grant No. 10874054, the NSAF of China under Grant No. 10676011, and the 2007 Cheung Kong Scholars Programme of China. We thank the Joint Supercomputer Center (Russian Academy of Sciences, Moscow) and CSCS (Manno) for providing supercomputer time. USPEX code is available on request from ARO.